\documentclass[aps,pre,reprint]{revtex4-2}

\usepackage{graphicx}
\usepackage{amsmath,amsfonts,amssymb}
\usepackage{orcidlink}
\usepackage{dcolumn}               
\usepackage{bm}                          
\usepackage{hyperref}

\begin{document}

\title{Mechanical waveform memory in an athermal random medium}
\author{Eamon Dwight\orcidlink{0009-0001-1137-5989}}
\author{D. Candela\orcidlink{0000-0002-5347-8361}}
\email[]{candela@physics.umass.edu}
\affiliation{Physics Department, University of Massachusetts, Amherst MA 01003}

\date{January 27, 2026}

\begin{abstract}
	Using numerical simulations it is shown that a random, athermal pack of soft frictional grains will store an arbitrary waveform that is applied as a small time-dependent shear while the system is slowly compressed.
	When the system is decompressed at a later time, an approximation of the input waveform is recalled in time-reversed order as shear stresses on the system boundaries.
	It is shown that this effect depends on friction between the grains, and is independent of some aspects of the friction  model.
	By systematically increasing the complexity of the stored waveform, it is found that a pack of $10^4$ grains can recall any one of 128 different waveforms with 100\% classification accuracy and 512 different waveforms with over 90\% classification accuracy, as measured by a neural net trained only on the inputs.
	This type of waveform memory might be observable in other types of athermal random media that form internal contacts when compressed such as crumpled sheets and nest-like fiber assemblies.
\end{abstract}

\maketitle

 \begin{figure}
 \includegraphics[width=\linewidth]{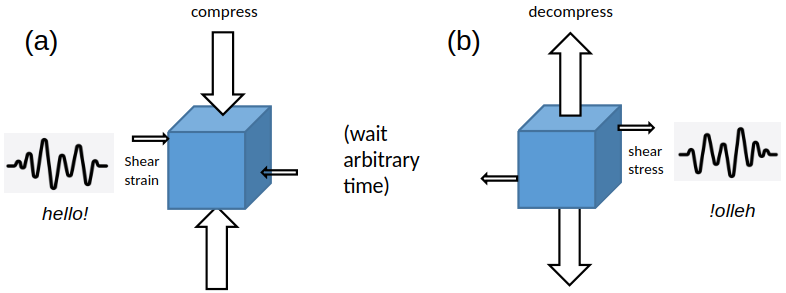}
 \caption{\label{fig-compressmem}Scheme for storage and recall of waveform data by a complex medium (blue cube) using compression as the reference input.
 	(a) The waveform data to be stored are applied as a small, time-dependent shear strain while the medium is progressively compressed.
 	Shear is used for data input as it is distinguished by spatial symmetry from the reference input (compression).
 	For a memory to be formed, a microscopic nonlinear interaction between compression and shear should form an imprint within the medium.
 	To retain the information the medium is held compressed for an arbitrary period.
 	(b) To read out the memory the medium is progressively decompressed, which effectively reapplies the reference input in reverse order.
 	If the medium has suitable properties, the waveform data are recalled in reversed order as shear stress at the system boundaries.}
 \end{figure}

\section{\label{intro}Introduction}
	Many types of condensed matter exhibit memory, i.e. the recall at a later time of inputs applied to the system at an earlier time~\cite{keim19, nagel23, barrat24, paulsen25}.
	Here we specialize to athermal systems, in which thermally-activated motion of the variables is much smaller than motion driven by boundary or body forces.
	Examples of athermal systems that show mechanical memory include a charge density wave model~\cite{povinelli99},  suspensions~\cite{pine05, corte08, keim11, keim13, paulsen14} and model glasses~\cite{fiocco14, mukherji19,keim20, lindeman25a, mungan25} composed of larger-than-colloidal particles, dry granular media like sand~\cite{toiya04, mueggenburg05, ren13, prados14, bandi18, zhao19, kramar21, lindeman21, ghosh22,  zhao22}, crumpled, corrugated, or twisted sheets~\cite{matan02, lahini17, vanbruggen19, bense21,  shohat22, dawadi24}, nest-like fiber assemblies~\cite{bhosale22}, and fabrics~\cite{crassous24}.
		
	Many such systems can recall a single number such as the amplitude of an applied shear~\cite{toiya04, mueggenburg05,ren13, vanbruggen19, zhao22}.
	For the CDW model and sheared suspensions it was found that the addition of noise enabled the recall of multiple driving values -- a discrete set of numbers~\cite{povinelli99, keim11, paulsen14}.
	
	More recently, it was found that densely coupled systems like model glasses, granular packs, and crumpled sheets can store and recall multiple discrete drive values without the addition of noise.
	This is typically attributed to hysteretic variable rearrangements termed hysterons, which enable return-point memory even when non-interacting~\cite{fiocco14, mukherji19, keim20, bense21, shohat22, lindeman25a}.
	More complex behaviors such as the ability to recall multiple discrete drive values without alternating driving are possible when the hysterons interact with each other~\cite{lindeman25, paulsen25}.

	It is natural to ask whether this sequence of progressively more complex memories can be extended to a dense set of recalled numbers, i.e. a function or a waveform.
	In this work we explore a mechanism for static waveform memory in an athermal granular pack.
	
	Some schemes for static analog storage of complex information like holography use nonlinear interactions within a medium to imprint correlations between a reference input and the information to be stored.
	Later, the reference is reapplied to the medium to recall the stored information.
	For a solid-like medium such as a random grain pack, we can envisage using compression as the reference and shear strain and stress as the input and output respectively for stored information (Fig.~\ref{fig-compressmem}).

	Specializing to a jammed, random pack of dry grains as the storage medium, friction between the grains could provide a grain-scale interaction between compression and shear that imprints and recalls the waveform data.
	Many of the unusual mechanical properties of granular media  are due to the network of contacts between grains~\cite{ohern03,somfai05,liu10}.
	In a large \emph{amorphous} granular pack additional contacts are formed continuously as the sample is compressed beyond the jamming point, and there are always contacts close to forming or separating~\cite{makse00,agnolin07b}.
	If there is friction at grain-grain contacts the transverse contact forces depend upon the the path by which the system reaches a given set of grain positions, creating possibilities for memory effects~\cite{mindlin53,elata96,johnson97}.

	In this work we use discrete-element method (DEM) simulations~\cite{cundall79, poschel05, matuttis14} to show that a random pack of frictional grains does indeed exhibit waveform memory of the sort illustrated in Fig.~\ref{fig-compressmem}.
	Similar waveform memory should be observable in other systems that form progressively more internal contacts when compressed such as fiber nests~\cite{bhosale22, gey25}, fiber bundles and yarns~\cite{panaitescu18, seguin22, dawadi25}, textiles~\cite{duhovic06, poincloux18, crassous24, singal24, gonzalez25}, crumpled sheets~\cite{matan02, cambou11, lahini17, vanbruggen19}, and possibly in other types of material with complex internal structures.
	Some of the results shown here were reported in an earlier short publication~\cite{candela23}.
	
	In common with the hysteron-based memory that has been observed recently in several other systems ~\cite{keim20, bense21, shohat22, lindeman25, lindeman25a, paulsen25} the waveform memory described here depends on hysteretic internal variables -- in this case frictional contacts between grains.
	Sec.~\ref{relationtohysteron} discusses how these different types of memory may be related.

\section{\label{demonstration}Demonstration of granular waveform memory}
	This section shows the results of simulations done with a small set of input waveforms to show that the waveform memory effect sketched in Fig.~\ref{fig-compressmem} actually occurs in simulated granular media, and to test its dependence on factors like the friction coefficient and friction model, sample size, and grain shape.
	Sec.~\ref{limits} shows results using more extensive sets of up to 1,024 different input waveforms  to probe the limits of memory complexity.

 \begin{figure}
 \includegraphics[width=\linewidth]{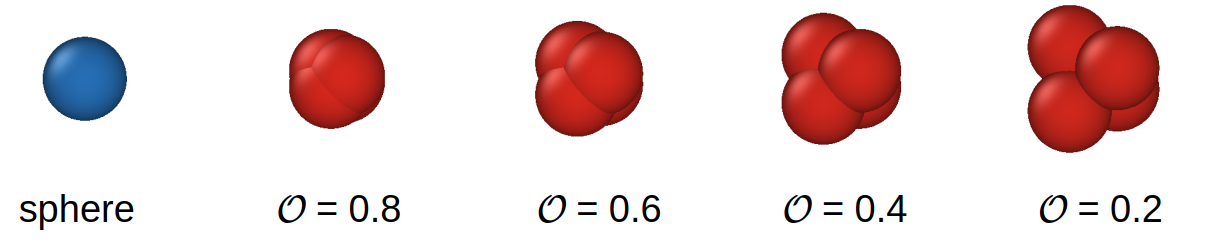}
 \caption{\label{fig-tetras}Examples of a single grain used in the simulations, which is either spherical (grain on left) or has the exterior shape of four partially overlapping spherical grainlets with their centers at the vertices of a rigid tetrahedron (remaining grains).
 	For tetrahedral grains the grainlet overlap $\mathcal{O}$ sets how non-spherical the grains are.
 	Most of the simulations reported here used tetrahedral grains with $\mathcal{O}=0.6$.}
 \end{figure}

\subsection{\label{sim-grains}Simulations of soft non-spherical grains}
	Most of the simulations shown here were carried out using non-spherical grains, with each grain modeled as four partially overlapping  spheres ~\cite{gallas93,nguyen19}  of radius $R$ fixed in a tetrahedral arrangement with a dimensionless overlap $\mathcal{O}$  defined to be zero when the spheres barely touch, and one when the spheres completely overlap (Fig.~\ref{fig-tetras}).
	As shown below, the memory effect was somewhat stronger for nonspherical grains than for spheres.
	Hence nonspherical grains with $\mathcal{O}=0.6$ were used except as noted.
 
\begin{figure}
 \includegraphics[width=\linewidth]{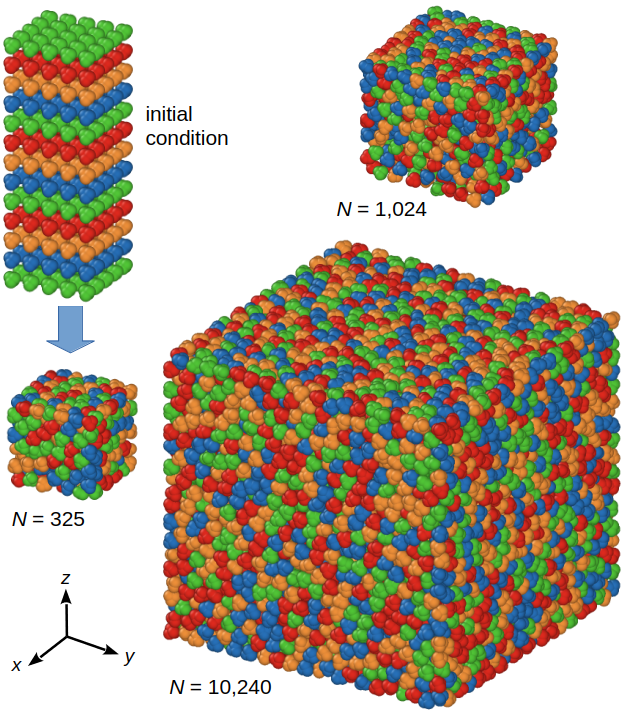}
 \caption{\label{fig-packs}Random packs of tetrahedral grains with $\mathcal{O}=0.6$ and various numbers of grains $N$, formed using the protocol described in the text.
 	The grains are confined by frictionless walls that are not shown.
 	For visibility the grains are rendered in four different colors but all grains have identical properties.
 	The initial condition for a 325-grain simulation is also shown, with the grains positioned on a regular lattice but with random velocities in a tall, rectangular box.
 	After evolving as a granular gas to randomize positions and orientations the samples are compressed in the $z$-direction to form the dense packs shown here.}
 \end{figure}

	A single friction coefficient $\mu$ was used for grain-grain contacts, rather than separate static and dynamic coefficients.
	The simulated granular packs (Fig.~\ref{fig-packs}) were subject to compressions with up to several percent volume reduction of the pack.
	In physical systems this would only be possible for grains made of a soft material such as an elastomer, so simulation parameters corresponding to millimeter-scale silicone rubber grains were used ($R=0.5$~mm, density $\rho = 1.2\times10^3$~kg/m$^3$, Young modulus $E=1.0\times10^7$~Pa, Poisson ratio $\nu=0.49$, and friction coefficient $\mu = 1.0$ except as noted).
	The only characteristic time that can be formed from these parameters is $t_c = R(\rho/E)^{1/2}$, which turns out to be about one-tenth the period of the highest-frequency vibrational mode of a moderately-compressed pack of these grains.
	Experimental time scales such as the time over which the sample is compressed must be compared with $t_c$, thereby setting the inertial number~\cite{dacruz05} for the grain motion.

	Contact forces between the grains were modeled by a Hertzian repulsive normal force along with three successively more realistic friction force models denoted H, M1, M2, detailed in Sec.~\ref{depend-friction}.
	Apart from model M2 these models have commonly been used to simulate granular media.
	Conventional DEM methods~\cite{cundall79, silbert01, poschel05, luding08, matuttis14, lammps25} were used to integrate the equations of motion for $N$ grains with these forces, confined by frictionless walls (Appendix~\ref{app-numerical}).

\subsection{\label{form-sample}Formation of dense granular samples.}
	For each sample size ($N \sim 300\dots10^5$), set of grain properties, and compression amplitude used, an initial simulation was used to prepare the packed granular sample.
	The grains were initially placed on a low-density cubical lattice with random velocities, and allowed to evolve as a granular gas  in a rectangular box with fixed walls to randomize grain positions and orientations~\cite{agnolin07a}.
	Then with the $+z$ wall  free to move an external pressure $p_z = (2\times10^{-3})E$ was used compress the grains into a dense pack.
	The initial box dimensions were chosen larger in the $z$ direction to give an approximately cubical pack of grains after this  $z$-compression, Fig.~\ref{fig-packs}.

	After the system came nearly to rest the $\pm z$ walls were fixed in place while the $\pm x, \pm y$ walls were used to compress and decompress the sample two times mimicking the compression cycles used later in the experiment simulations.
	 The grain-grain friction coefficient $\mu$  was set to zero during the entire sample preparation simulation.
	 This ``lubricated assembly'' protocol gives a numerically inexpensive way of preparing a well-annealed sample of frictional grains, with most grain-grain contacts well below the eventual Coulomb  sliding threshold~\cite{makse99,thornton00,agnolin07a}.

 \begin{figure}
 \includegraphics[width=\linewidth]{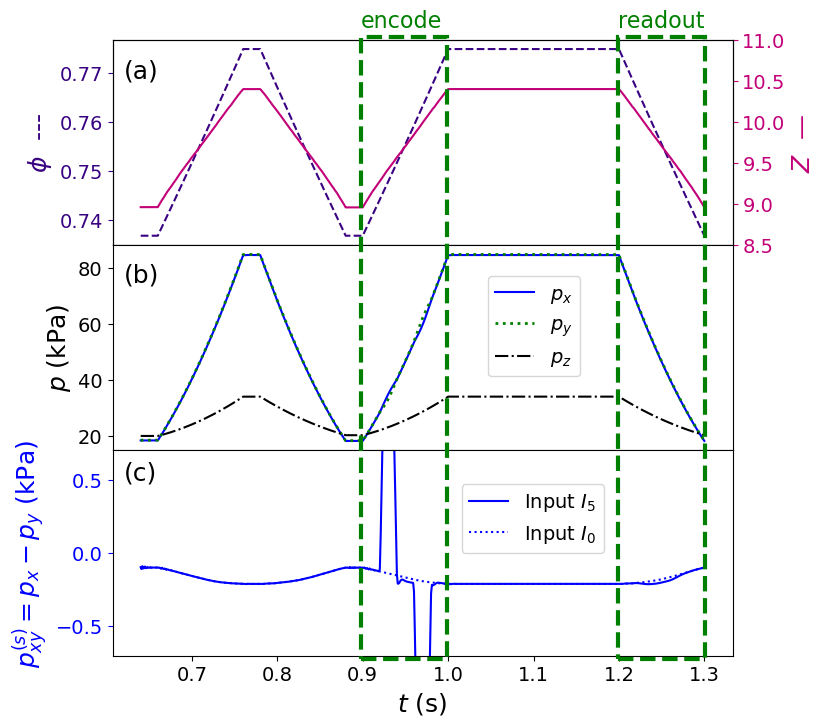}
  \caption{\label{fig-thingsvst} Various quantities versus time $t$ for a memory-experiment simulation.
 	The sample preparation period with zero friction ($t<0.64$\,s) is not shown; conversely friction is active for the entire period shown here.
	After a final preparatory compression cycle there is an encoding period (left dashed box) during which the input signal $\gamma_0 I_5(u)$ (Fig.~\ref{fig-inputs}(a)) is applied as an $(x-y)$ pure shear strain while the system is compressed in the $x$-$y$ plane.
	The system is held compressed during a storage period ($t=1.0$ - 1.2~s), then decompressed (right dashed box) to read out the memory response as shear stress $p_x-p_y$ on the system boundaries.
 	(a)~Sample filling factor $\phi(t)$ and average coordination number $Z(t)$.
 	Note tetrahedral particles typically pack more densely than spheres~\cite{hajiakbari09}.
 	(b)~Measured pressures on the $x$, $y$, and $z$ walls.
 	The responses of $p_x,p_y$ to the input signal are barely visible during the encoding period.
 	(c)~Difference between $p_x$ and $p_y$ on an expanded vertical scale, shown both when input $I_5(u)$ is applied and when the zero-shear input $I_0(u)$ is applied.
 	On this expanded scale the response to $I_5(u)$ during the encoding period is readily visible, while differences during the readout period between the responses to $I_5(u)$ and $I_0(u)$ are barely visible.
 	These differences, shown expanded in Fig.~\ref{fig-memeffect}, constitute the waveform memory effect.}
 \end{figure}

 \begin{figure}
 \includegraphics[width=\linewidth]{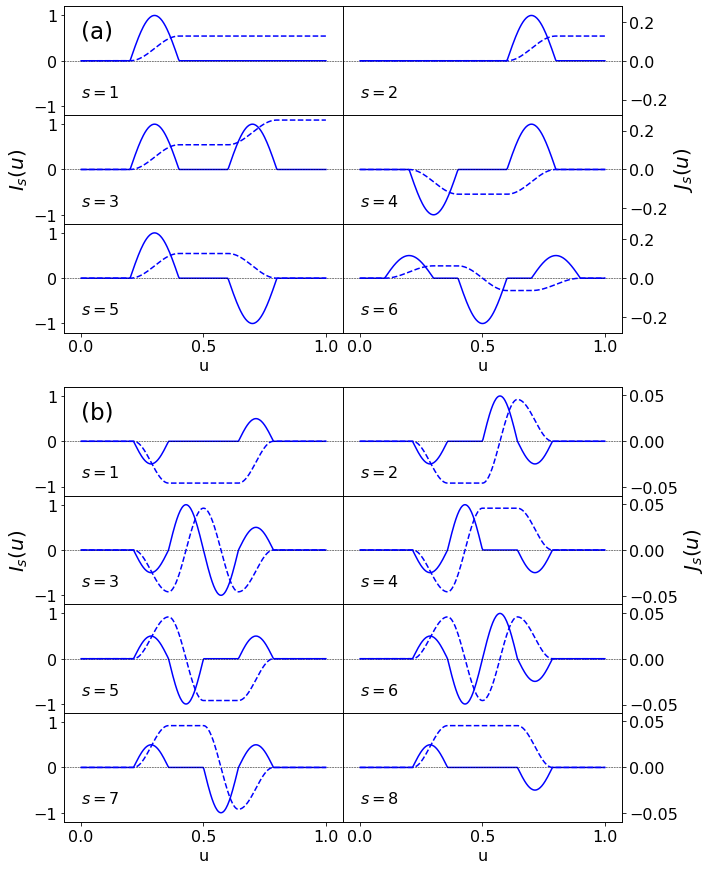}
 \caption{\label{fig-inputs}Waveforms used during compression as inputs for memory experiments.
 	In these figures the solid curves show $I_s(u)$ giving the applied shear strain as a function of the dimensionless compression $u$ for the $s^\textit{th}$ input waveform.
 	In all cases $I_s(u)$ is the sum of one or more half-cycle cosine bumps (Appendix~\ref{app-inputsigs}).
 	It was found (Fig.~\ref{fig-memeffect}) that the imprinted and recalled memory was proportional not to $I_s(u)$ but rather to its integral $J_s(u) = \int_0^u I_s(v)dv$ which is shown as the dashed curves.
 	(a)~Set of $S=6$ input signals used in Sec.~\ref{demonstration} for a demonstration of the memory effect and to test its dependence on friction and other simulation parameters.
 	(b)~Example for $L=3$ of the set of $S=2^L$ binary-word input signals used in Sec.~\ref{limits} to test the limits of memory complexity.
 	For these signals $J_s(u)$ follows the bit pattern of the $L$-bit binary representation of $s-1$ (Appendix~\ref{app-inputsigs}).
 	For example, $J_2(u)$ (dashed curve in upper right of (b)) is $-,-,+$ to follow the 3-bit binary representation 001 of $s-1=1$.}
 \end{figure}

\subsection{\label{mem-expts}Memory experiments.}
	 Each  sample prepared as above was used as the starting point for multiple  memory-experiment simulations using different shear input waveforms.
	 For these simulations $\mu$ was set to the chosen value (1.0 except as noted).
	To exhibit the memory effect the grain pack was slowly and linearly compressed while simultaneously applying an arbitrary input waveform as a small shear strain, similar to the sketch of Fig.~\ref{fig-compressmem}(a).
		The compression was applied by moving the four walls $\pm x$, $\pm y$ inward simultaneously and linearly in time over a period $t_0 = (1.8\times 10^4)t_c$, to  change the sample volume by $\Delta V/V = -\delta_0$.
	With $\delta_0=0.05$ (used except as noted), the average number of contacts on a grain $Z$ was measured to increase from 9.0 to 10.4 as the sample was compressed (Fig.~\ref{fig-thingsvst}(a)), remaining between the frictional ($Z=4$) and frictionless ($Z=12$) isostatic values~\cite{donev07,henkes10},\footnote{A rough explanation for the variation of $Z$ with compression is as follows: When the system is first jammed (at infinitesimal pressure) $Z$ should equal the frictional isostatic value. But as the system is further compressed more and more frictional contacts are mobilized,  pushing $Z$ towards the larger frictionless isostatic value.}.
	The inertial number $I = \dot{\epsilon}\sqrt{m/Dp}$ (with $\dot{\epsilon}$ the strain rate, $m,D$ the grain mass and diameter, and $p$ the pressure) was always less  than $4\times10^{-5}$, giving a nearly quasistatic compression~\cite{dacruz05,agnolin07b}.	
	
	The compression was parameterized by a dimensionless variable $u(t)$ that went from zero to one as the sample was compressed, then back to zero when the sample was decompressed.	
	The sample was compressed and decompressed once with friction but without applied shear (Fig.~\ref{fig-thingsvst}, left side).
	Then, during the course of a final compression, an input was applied by small additional movements of the $\pm x$ walls inward while moving the $\pm y$ walls outward (or vice versa), so as to create a pure shear of the sample $\gamma(u) = \gamma_0 I_s(u)$.
	Here $\gamma_0 $ (typically $10^{-3}$) set the scale of the shear strain and  $I_s(u), s=0\dots S$ were memory input waveforms with $-1\leq I_s(u) \leq 1$ used for $S+1$ separate experiment simulations (Fig.~\ref{fig-inputs})

	Each set of input waveforms included the zero signal signal $I_0(u)=0$, along with $S$ non-zero input waveforms $I_1(u)\dots I_S(u)$.
	This section shows results using the $S=6$ nonzero shear input waveforms  shown in Fig.~\ref{fig-inputs}(a).
	These waveforms are sums of 1-3 half-cycle cosine bumps, chosen somewhat arbitrarily to test if a true waveform memory is present.
	Specifically $I_1(u), I_2(u)$ test whether the response can be localized to different $u$'s, while $I_3(u)\dots I_6(u)$ test whether sums of multiple inputs localized at different $u$'s give a correspondingly summed outputs.
	A more systematically varying set of input waveforms (Fig.~\ref{fig-inputs}(b)) is used in Sec.~\ref{limits} below. 

	Except as otherwise noted, the following parameters were used for the memory-experiment simulations in this paper:
	Grain shape tetrahedral with overlap $\mathcal{O}=0.6$, number of grains $N=$~10,240, grain-grain friction coefficient  $\mu=1.0$, transverse force law Hookean (model H, Sec.~\ref{depend-friction}), compression amplitude $\delta_0 = 0.05$, and input-signal strain amplitude $\gamma_0 = 10^{-3}$.
	In Secs. ~\ref{depend-friction} and \ref{depend-shape} below each of these parameters is varied individually while keeping the other parameters at these standard values.

 \begin{figure}
 \ \includegraphics[width=\linewidth]{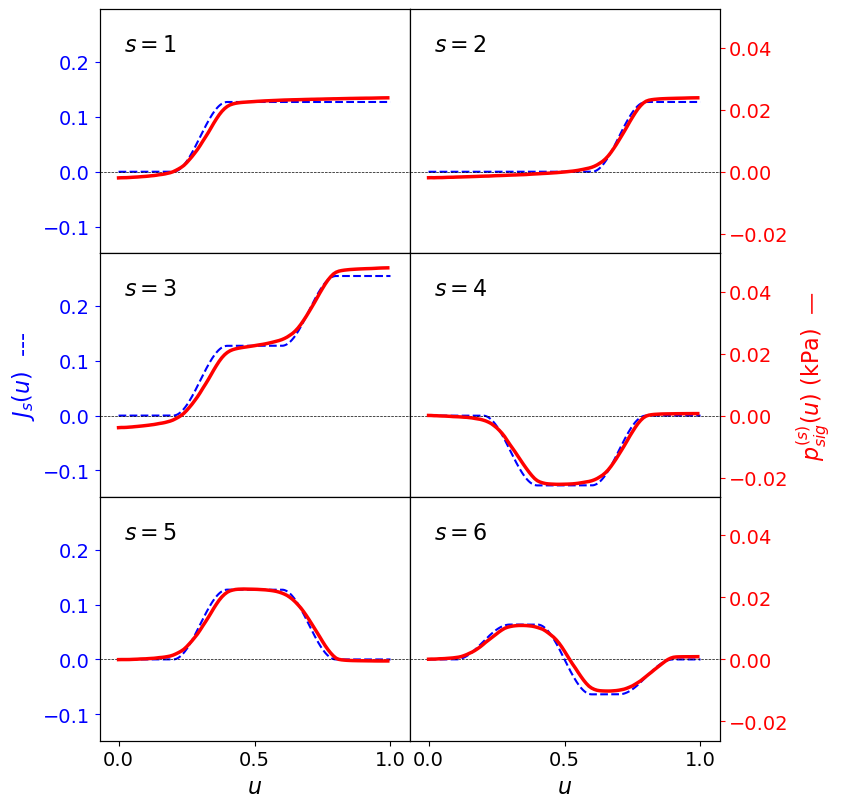}
 \caption{\label{fig-memeffect} Recalled shear pressure signal $p_\mathit{sig}^{(s)}(u) = -(p_{xy}^{(s)} - p_{xy}^{(0)})$ when the sample was decompressed, for the six input signals $s=1\dots6$ shown in Fig.~\ref{fig-inputs}(a) (solid lines).
 	For comparison the dashed lines show the corresponding  integrated shear inputs $J_s(u)$ applied earlier when the sample was compressed.
 	A single gain factor $\mathcal{G}$ was computed to minimize the least-squares difference between $\mathcal{G}E\delta_0\gamma_0 J_s(u)$ and $p_\mathit{sig}^{(s)}(u)$ summed over all six signals, and used to scale all six  plots equally.}
 \end{figure}

 \begin{figure}
 \ \includegraphics[width=\linewidth]{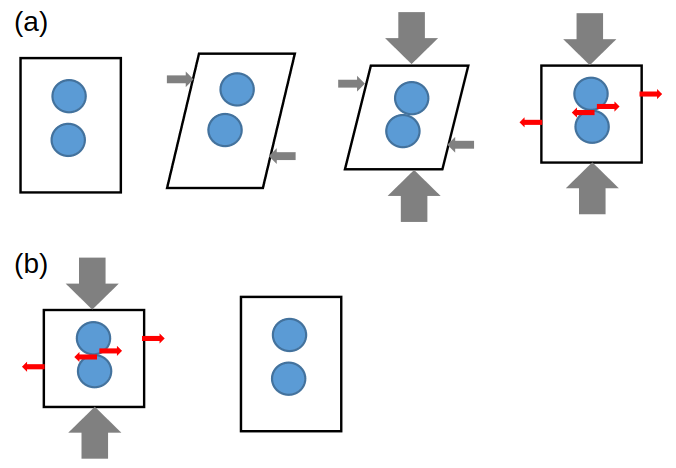}
 \caption{\label{fig-explanation}Proposed explanation for waveform memory as sketched in Fig.~\ref{fig-compressmem} for a random pack of frictional grains.
 	Here the black parallelogram represents the overall pack, while the blue circles represent two grains within the pack that come into and out of contact at a certain degree of compression $u_0 < 1$.
 	For this cartoon the input signal is shown as a simple shear of the pack rather than a pure shear as used in the simulations.
 	(a)~Imprinting memory during compression: If the pack is macroscopically sheared to some degree $
 	\gamma_0 I(u_0)$ at the point when the grains come into contact, then after the shear is removed the contact will be in a state of transverse stress.
 	The contact stress will be transmitted in some small degree to the system boundaries, as a boundary shear stress opposite $\gamma_0 I(u_0)$ (upper right).
 	Summing over all contact forces, at the end of the compression ($u=1$) the shear stress measured at the boundaries will be $p_{xy}(1) = -C\gamma_0 \int_0^1I(u)du$ for some small constant $C$ with $\gamma_0 I(u)$ the shear strain waveform applied during compression.
 	(b)~Readout of the memory during decompression: Assuming this contact comes apart during decompression at the same point $u_0$ at which it was formed during compression, the transverse stress in the contact (and transmitted to the boundaries) is relieved at $u=u_0$.
 	Summing over all contacts, during decompression $p_{xy}(u) = p_{xy}(1) - \left(-C\gamma_0 \int_u^1 I(v)dv\right) = -C\gamma_0 \int_0^u I(v)dv$.}
 \end{figure}

\subsection{\label{existence}{Existence of the memory effect.}}
	Figure~\ref{fig-thingsvst}(b,c) shows the pressures $p_x$, $p_y$, $p_z$ measured on the sample walls during a typical simulation run when shear input signal $\gamma_0I_5(u)$ is applied, and the resulting measured boundary shear stress $p_{xy}^{(5)} = p_x - p_y$.
	While the response of $p_{xy}^{(5)}$ to the input signal $I_5(u)$ during compression is clearly visible, the recalled response when the sample is later decompressed is barely visible due to the background signal observed even when the zero input $\gamma_0 I_0(u)$ is used (Fig.~\ref{fig-thingsvst}(c)).
	This background reflects the slight $x$-$y$ asymmetry of any specific realization of the random pack like those shown in Fig.~\ref{fig-packs}.

	In Fig.~\ref{fig-memeffect} the zero-input background (measured in the separate simulation using $\gamma_0 I_0(u)$) is subtracted to give the processed response signal $p_\textit{sig}^{(s)} = -(p_{xy}^{(s)} - p_{xy}^{(0)})$, with the minus sign used to simplify equations below.
	It can be seen that $p_\textit{sig}^{(s)}(u)$ is nearly proportional to the integrated input signal $J_s(u)$, demonstrating a robust but imperfect waveform memory effect:
\begin{equation}\label{psig}
	p_\mathit{sig}^{(s)}(u) \approx \textit{const}\times\int_0^u  \gamma_0 I_s(v)dv \equiv \textit{const}\times \gamma_0 J_s(u).
\end{equation}
	While it was initially unclear to us why the integrated shear $\gamma_0 J_s(u)$ should be recalled rather than the shear itself $\gamma_0 I_s(u)$, this agrees with a possible explanation for the memory effect illustrated in Fig.~\ref{fig-explanation}.

	As precise forces on all grains are known in simulations like this, it might be thought the the memory mechanism proposed in Fig.~\ref{fig-explanation} could be tested directly by examining the statistics of friction forces.
	But it appears that the ``trace'' of the memory consists of tiny modifications to the friction forces on a much larger random background, and we have not found direct evidence of the stored waveforms in the contact force distributions.

	During the storage period between encoding and recalling the signals (1.0-1.2~s in Fig.~\ref{fig-thingsvst}) the maximum grain velocity was found to decay exponentially towards the numerical noise floor, implying that the signals could be stored indefinitely.
	However these simulations do not include creep of contact forces, which would likely limit storage times in some physical systems~\cite{dijksman22, farain24, korchinski25, jules20}.

\subsection{Gain and fidelity.}
	In the quasistatic limit the only parameter of the simulations with dimensions of $p_\mathit{sig}^{(s)}(u)$ (i.e. pressure) is the grain-material Young modulus $E$, so based on Eq.~\ref{psig} we introduce a dimensionless ``gain'' parameter $\mathcal{G}$ such that
\begin{equation}\label{ggdef}
	p_\mathit{sig}^{(s)}(u) \approx \mathcal{G}E\delta_0\gamma_0 J_s(u).
\end{equation}
	As the compression amplitude $\delta_0$ controls the number of contacts formed, the recalled signal $p_\mathit{sig}^{(s)}(u)$ might be expected to be proportional to $\delta_0$ as implied by Eq.~\ref{ggdef}.
	This is tested below by measuring the actual dependence of $\mathcal{G}$ on $\delta_0$, expected to be weak.

	As detailed in Appendix~\ref{app-dotprod}  $\mathcal{G}$ is determined from the simulation results  as the value that minimizes the error in Eq.~\ref{ggdef} summed over the $s=1\dots S$ signals in the set.
	Then a ``fidelity'' $\mathcal{F} \in [0,1]$ is computed with $\mathcal{F}=1$ indicating a perfect agreement between the recalled signals scaled by the fitted $\mathcal{G}$ and the input signals.

	For the simulation results shown in Fig.~\ref{fig-memeffect} this gives $\mathcal{G} = 0.354$, $\mathcal{F} = 0.910$.	
	While it is interesting that $\mathcal{G} \sim \mathcal{O}(1)$,  for a theoretical calculation of $\mathcal{G}$ it would be necessary to connect the macroscopic applied strain $\gamma_0 I_s(u)$ to the distribution of transverse grain movements at contacts, and similarly to connect grain-scale friction forces to the macroscopic wall stress $p_{xy}^{(s)}(u)$.
	These connections are nontrivial due to non-affine grain motion and the creation of contacts as the sample is compressed \cite{makse99, agnolin07b, wilmarth25}.

 \begin{figure}
 \ \includegraphics[width=\linewidth]{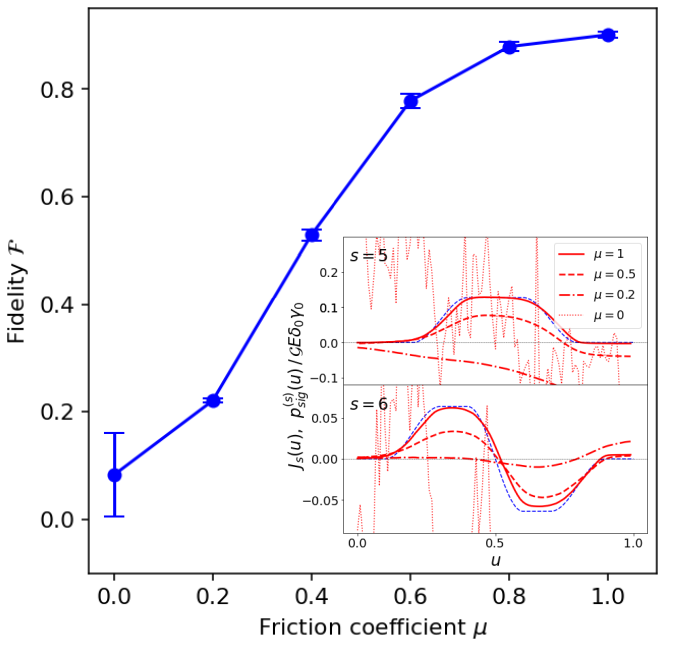}
 \caption{\label{fig-varymu} Effect of grain-grain friction coefficient $\mu$ on the memory fidelity $\mathcal{F}$, for the set of six input waveforms shown in Figs.~\ref{fig-inputs}(a) and Fig.~\ref{fig-memeffect}.
 	Here and below error bars were determined by running the entire set of simulations several times with different realizations of the random grain pack.
 	Inset: Recalled waveform (red) vs input waveform (blue dashed) for two of the six inputs $J_5(u)$, $J_6(u)$ and various values of $\mu$.
 	(For each $\mu$ value the gain $\mathcal{G}$ was computed to minimize the error for all six input waveforms as in Fig.~\ref{fig-memeffect}, even though only two are shown here.)
 	The fit to the input waveform becomes progressively worse as $\mu$ is reduced, and when $\mu=0$ numerous particle rearrangements on compression make the recalled signal very noisy.}
 \end{figure}

 \begin{figure}
 \ \includegraphics[width=\linewidth]{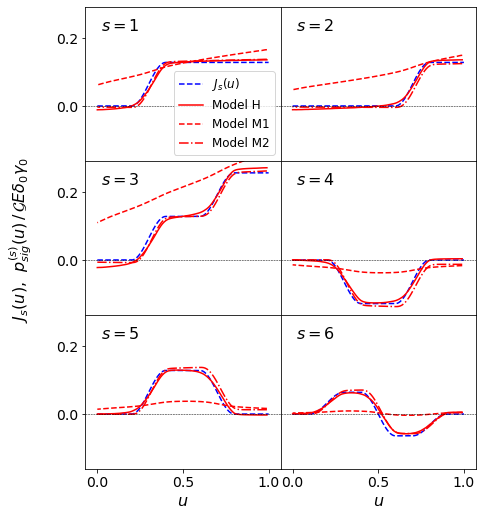}
 \caption{\label{fig-varymodel} Effect of varying the friction model (transverse force law).
 	For each model (H, M1, M2) the gain $\mathcal{G}$ was computed for the best joint fit of the recalled signals to the six input waveforms of Figs.~\ref{fig-inputs}(a).
	The simplest model H and the most realistic model for viscoelastic spheres M2 show similar memory effects (fidelities $\mathcal{F}=0.910, 0.891$ respectively), while the intermediate model M1 shows poor memory ($\mathcal{F}=0.460$).
	Apart from this figure model H was used for all simulation results shown in this paper.}
 \end{figure}

 \begin{figure}
 \includegraphics[width=\linewidth]{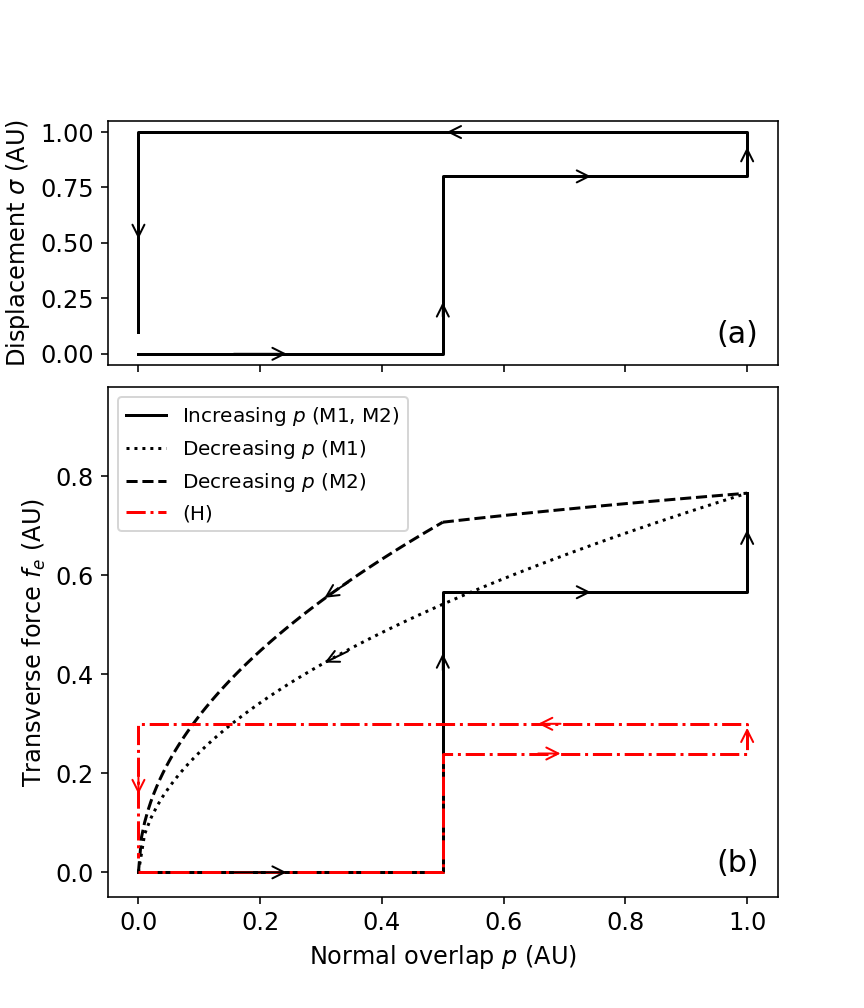}
 \caption{\label{fig-m1m2h}(a) Assumed path of transverse displacement $\sigma$ vs. normal overlap $p$ (both in arbitrary units) for two grains in contact, to illustrate the differences between transverse force laws M1, M2, and H.
 	After the grains are brought into contact and the overlap is increased to $p=0.5$, the transverse displacement is increased from $\sigma=0$ to $\sigma=0.8$.
 	Then the overlap is increased further to $p=1$ before the transverse displacement is increased further to $\sigma=1$.
 	Finally the overlap is reduced to $p=0$, taking the two grains out of contact and resetting the accumulated transverse displacement to $\sigma=0$.
 (b) Resulting elastic transverse force $f_e$ for the force laws M1, M2, and H.
 	The transverse force is the same for M1 and M2 as the overlap $p$ is increased.
 	But as $p$ is decreased to zero the force from M2 has a more complex variation with $p$ reflecting the application of transverse displacement $\sigma$ at two different overlaps $p=0.5$ and $p=1.0$.
 	Conversely M1 does not have this history information as it simply assumes $f_e\propto p^{1/2}$ for decreasing $p$.
 	Finally, the red dash-dot curve shows $f_e$ vs. $p$ for the Hookean transverse force law H (with a smaller prefactor, $k_H <k_M$, for clarity).
 	Like M1, H does not store complex history information -- rather $f_e(p)$ for H is directly proportional to $\sigma(p)$ in graph~(a).}
 \end{figure}

\subsection{\label{depend-friction}Dependence on friction coefficient and friction model}
	It is proposed  in Fig.~\ref{fig-explanation} that the waveform memory is due to grain-grain friction.
	To test this additional simulations were carried out varying the friction coefficient $\mu$ and the transverse force law used to implement friction.
	Figure~\ref{fig-varymu} shows the effect of varying $\mu$.
	Both the appearance of waveform memory and the calculated memory fidelity $\mathcal{F}$ decrease rapidly with decreasing $\mu$, with the memory effect disappearing completely in the frictionless limit $\mu\rightarrow0$.
 
	To check if the memory effect could be an artifact of the specific friction model used, simulations were carried out using three different transverse (friction) force laws.	
	A Hertzian repulsive normal force~\cite{brilliantov96,schwager08} was used for all three models,
\begin{equation}\label{fn}
	f_n = \mbox{max}(0, k_n p^{3/2} + \gamma_n p^{1/2}\dot{p})
\end{equation}
as appropriate for viscoelastic rather than plastically deforming grains \cite{walton93,vuquoc99,thornton13}.
	Here $p$ is the normal overlap of two grains, $\dot{p}$ its rate of change, $k_n=2^{-3/2}(4/3)R^{1/2}E/(1-\nu^2)$, and  $\gamma_n/k_n = (0.23)t_c$ was used~\footnote{This relatively large (low-$Q$) damping constant $\gamma_n$ for grain-grain forces was chosen for numerical efficiency -- to enable a larger simulation time step to be used without numerical instability -- rather than modeling viscoelastic forces in any particular physical material.}.
	There is significant rolling motion at contacts when a granular medium is compressed~\cite{kuhn04,benson22}; for this work the rolling and twisting resistances were set to zero.

	The three friction models tried H, M1, M2 all include transverse elastic and damping force vectors $\mathbf{f}_e$, $\mathbf{f}_d$ with the magnitude of the total transverse force limited by the Coulomb criterion $|\mathbf{f}_e + \mathbf{f}_d| < \mu f_n$ using the algorithm of
 Ref.~\cite{luding08}.
	
	Model H (``Hooke''), used for most of the results shown in this paper, has a linear transverse spring and dashpot
\begin{equation}\label{fth}
	\Delta\mathbf{f}_e = k_H\Delta \bm{\sigma},\ \ \ \mathbf{f}_d = \gamma_H \dot{\bm{\sigma}}
\end{equation}
with  $\bm{\sigma}$ the vector relative sliding motion between the two grain surfaces and $\dot{\bm{\sigma}}$ its rate of change.
	Apart from the damping term this is the original friction model of Cundall and Strack~\cite{cundall79}.
	Here $k_H$ was set to a typical inverse transverse compliance from model M1 below~\cite{johnson85}, and $\gamma_H$ was set using $\gamma_H/k_H = (0.3)\gamma_n/k_n$ to give similar damping for normal and transverse contact motion.

	Model M1 (``Mindlin-1'') improves upon model H by making the transverse compliance dependent upon the normal overlap $p$, using a linearized, no-slip version~\cite{johnson97},\footnote{As can be seen from Eq.~\ref{ftm1}, the model of ref.~\cite{johnson97} is linearized in the transverse displacement  $\Delta\bm{\sigma}$ but it retains the Hertzian nonlinearity in the normal overlap $p$.} of the transverse force calculated by Mindlin and Deresiewicz for contacting elastic spheres \cite{mindlin49,mindlin53}.
	In this model the elastic force is accumulated using
\begin{equation}\label{ftm1}
	\Delta\mathbf{f}_e = k_Mp^{1/2} \Delta\bm{\sigma},\ \ \ \mathbf{f}_d = \gamma_M p^{1/2}\dot{\bm{\sigma}}
\end{equation}
	with $k_M = 3k_n(1-\nu)/(2-\nu)$ ~\cite{johnson85, johnson97, makse99}.
	As with model~H, $\gamma_M/k_M = (0.3)\gamma_n/k_n$ was used.

	According to Eq.~\ref{ftm1} the transverse spring constant $k_Mp^{1/2}$ and hence the stored elastic energy varies with $p$, which would allow the unphysical generation of elastic energy by decreasing $p$ at fixed transverse displacement $\bm{\sigma}$~\cite{elata96, johnson97}.
	To avoid this, an approximation due to Walton is typically used \cite{agnolin07a,thornton13}: the elastic force $\mathbf{f}_e$ is reduced proportionally to $p^{1/2}$ when $p$ decreases.
	As discussed in Ref.~\cite{elata96} although rescaling $\mathbf{f}_e \propto p^{1/2}$ can prevent the spurious generation of elastic energy, this generally \emph{overestimates} the energy loss in the linearized Mindlin model.

	Model M1 has frequently been used for DEM simulations of frictional granular matter~\cite{makse99,makse04,agnolin07a,thornton13},\footnote{Model M1 corresponds to the LAMMPS transverse force law \texttt{tangential mindlin\_rescale/force}, see App.~\ref{lammps} }.
	Interestingly, when the nominally more realistic model M1 is substituted for model H, the memory effect reported here nearly disappears, Fig.~\ref{fig-varymodel}.
	We find this is due to the approximate handling in model M1 of the reduction of $\mathbf{f}_e$ on decreasing $p$.	
	
	Model M2 (``Mindlin 2'') avoids this approximation by accurately computing the change of $\mathbf{f}_e$ with decreasing overlap $p$ given by the linearized Mindlin model of Ref.~\cite{johnson97}.
	One way to do this (used here, see App.~\ref{app-m2} for a numerical implementation) is to represent the transverse elastic force in a contact as an integral over contributions from different values of $q = p^{1/2}$, i.e. $\mathbf{f}_e = \int_0^\infty \mathbf{f}(q) dq$.
	The first of Eqs.~\ref{ftm1} is implemented by adding $k_M \Delta\bm{\sigma}$ to $\mathbf{f}(q)$ in the interval $0 < q < p^{1/2}$, and when $p$ decreases $\mathbf{f}(q)$ is set to zero for $q>p^{1/2}$.
	Note this is not  a new friction model, but rather a more accurate representation~\footnote{Unlike model M1, model M2 has the following property emphasized in Ref.~\cite{johnson97}: After the normal overlap $p$ is reduced from a larger value $p_2$ to a smaller value $p_1$, the transverse force is the same as it would have been following a similar displacement trajectory in which $p$ never exceeded $p_1$.  Physically this is because the ring-shaped area over which the spheres came into contact when $p > p_1$ is no longer in contact, hence no longer contributes to the transverse force.} than model M1 of the transverse force law of Ref.~\cite{johnson97}.
	Model M2 has not typically been used for DEM simulations because it requires storing a vector-valued function $\mathbf{f}(q)$ of history information for each contact.
	Thus model M2 is more complicated to implement and requires more computational resources than model M1.
	Figure~\ref{fig-m1m2h} illustrates the differences in the transverse elastic response of a single contact predicted by models H, M1 and M2.

	With the more-faithful linearized Mindlin model M2, the memory effect is very similar to that seen for the simplest model H, Fig.~\ref{fig-varymodel}.
	Although friction models other than the three considered here have been used \cite{walton86,shafer96,poschel05} models H and M2 are sufficiently different to suggest that the memory effect is insensitive to details of the transverse force law, provided it does not lose memory information as does model M1 (Fig.~\ref{fig-m1m2h}).
	
	In the Mindlin-Deresiewicz theory~\cite{mindlin53} and its linearized form (Ref.~\cite{johnson97} and model M2), each contact retains detailed history of the transverse strains  applied as the contact is compressed~\cite{johnson97}.
	To show that the waveform memory discussed here is an emergent property of a grain pack with many contacts rather than depending on the complex memory behavior of individual contacts, all of the simulation results shown in this paper apart from those shown in Fig.~\ref{fig-varymodel} used the linear model H for friction forces.

 \begin{figure}
 \ \includegraphics[width=\linewidth]{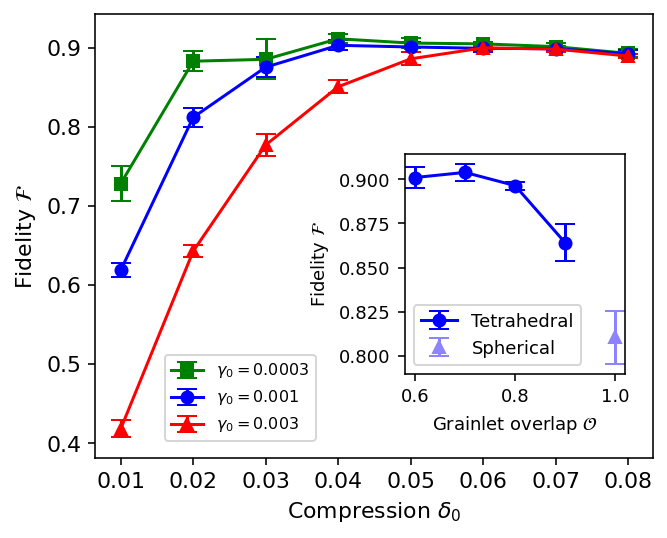}
 \caption{\label{fig-ffdelta0oo} Effect on the waveform-memory fidelity $\mathcal{F}$ of varying the amplitude $\delta_0$ of the volume compression used as a reference input, for three different values of the waveform-input shear amplitude $\gamma_0$.
 	The fidelity is good ($\mathcal{F} \approx 0.9$) over a range of compression amplitudes provided the waveform-input shear is much smaller than the compression, $\gamma_0 \leq 0.01\delta_0$. 
 	Inset:   Effect of grain shape on $\mathcal{F}$.
	Fidelity results are shown for tetrahedral grains with four different values of the overlap parameter $\mathcal{O}$, as well as for spherical grains (Fig.~\ref{fig-tetras}).
	The fidelity generally decreases as the grain shape becomes closer to spherical. }
 \end{figure}

 \begin{figure}
 \ \includegraphics[width=\linewidth]{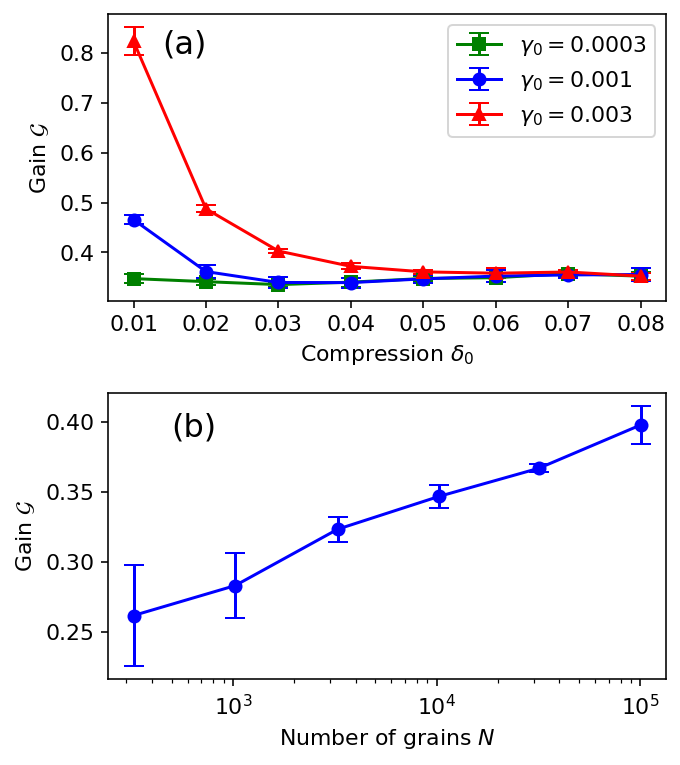}
 \caption{\label{fig-ggdelta0ng} (a) Variation of the dimensionless gain $\mathcal{G}$ of the waveform memory with compression amplitude~$\delta_0$.
 	Provided the input-waveform shear amplitude satisfies $\gamma_0 \ll \delta_0$, $\mathcal{G}$ is nearly independent of $\delta_0$.
 	(b) Variation of $\mathcal{G}$ with the number of grains $N$ in the granular pack.
 	The dependence is weak, approximately obeying $\mathcal{G} \propto N^{0.08}$.}
 \end{figure}

\subsection{\label{depend-shape}Dependence on grain shape and scaling with simulation parameters}
	The inset of Fig.~\ref{fig-ffdelta0oo} shows the effect of changing the grain shape on the memory fidelity.
	As the grains are made closer to spherical by reducing the sphere overlap from its default value ($\mathcal{O}=0.6$, Fig.~\ref{fig-tetras}) the fidelity $\mathcal{F}$ systematically decreases, although a significant memory effect ($\mathcal{F} =0.81$) is seen even for simulations of spherical grains.
	Using non-spherical grains reduces the number of irreversible grain rearrangements that occur when the sample is compressed.
	Similarly it was found in Ref.~\cite{kramar21} that large grain movements destroy a different type of granular memory when the medium is sheared.
	
	Recently Sun, et al.~\cite{sun24} used MRI experiments to show that small deviations from spherical grain shape had significant effects on contact statistics and orientational ordering in frictional grain packs.
	Conversely the memory effect described here appears to interpolate smoothly between spherical and slightly non-spherical grain shapes (Fig.~\ref{fig-ffdelta0oo} inset).

	Figure~\ref{fig-ffdelta0oo} also shows the effects on the fidelity $\mathcal{F}$ of changing the amplitude $\delta_0$ of the volume compression used as a reference input and the amplitude $\gamma_0$ of the shear strain used for memory inputs.
	The fidelity appears nearly independent of the compression $\delta_0$ provided the input-signal shear is much smaller than the compression, $\gamma_0 \ll \delta_0$.

	Figure~\ref{fig-ggdelta0ng} shows how the dimensionless gain $\mathcal{G} = p_\textit{sig}^{(s)}(u)/E\delta_0\gamma_0J_s(u)$ (see Eq.~\ref{ggdef}) varies with the compression amplitude $\delta_0$ and the number of grains in the sample $N$.
	Consistent with the contact-based mechanism proposed in Fig.~\ref{fig-explanation} the gain $\mathcal{G}$ is nearly independent of $\delta_0$ provided $\gamma_0 \ll \delta_0$.
	Also $\mathcal{G}$ is a weakly dependent on the sample size $N$, but it is not clear from these data whether $\mathcal{G}$ would  approach a constant as $N \rightarrow \infty$, i.e. whether $\mathcal{G}$ can be considered to be a bulk property of these grain packs.
	
 \begin{figure}
 \ \includegraphics[width=\linewidth]{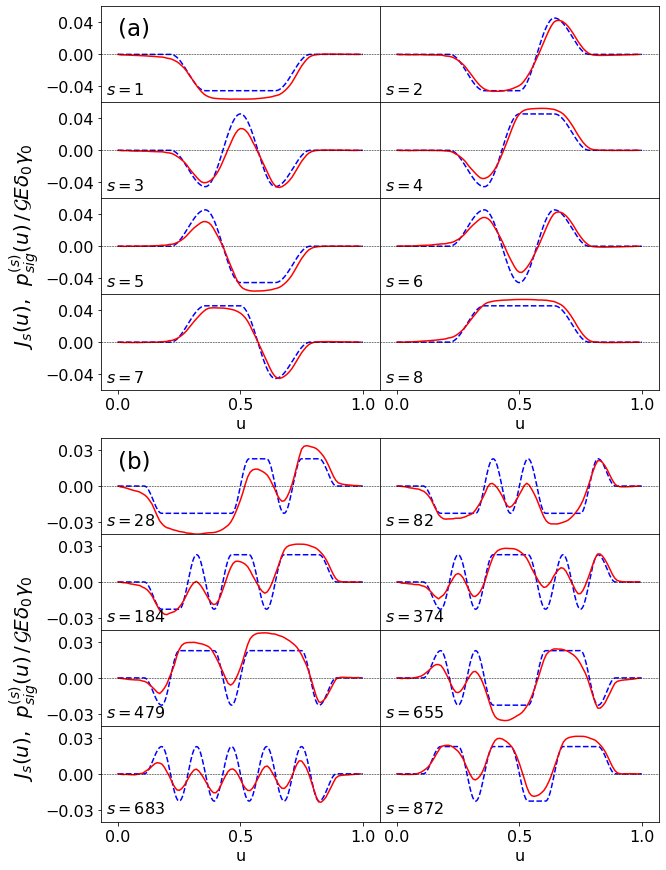}
 \caption{\label{fig-bwmem} Recalled memory signals $p_\textit{sig}^{(s)}(u)$ (solid red lines) compared with integrated input signals $J_s(u)$ (dashed blue curves) using the $L$-bit binary-word input signals for (a)~$L=3$ (Fig.~\ref{fig-inputs}(a)) and (b)~$L=10$.
 	As in Fig.~\ref{fig-memeffect} the gain $\mathcal{G}$ used to scale  $p_\textit{sig}^{(s)}(u)$ for each $L$ was fitted using all $S = 2^L$ signals.
 	In~(a) all eight $L=3$ signals are shown, while in~(b) eight of the 1,024 $L=10$ signals are shown.
 	The signals were selected randomly except that one of the most rapidly varying signals ($s=683$, representing binary 1010101010) was purposely included.}
 \end{figure}

 \begin{figure}
 \ \includegraphics[width=\linewidth]{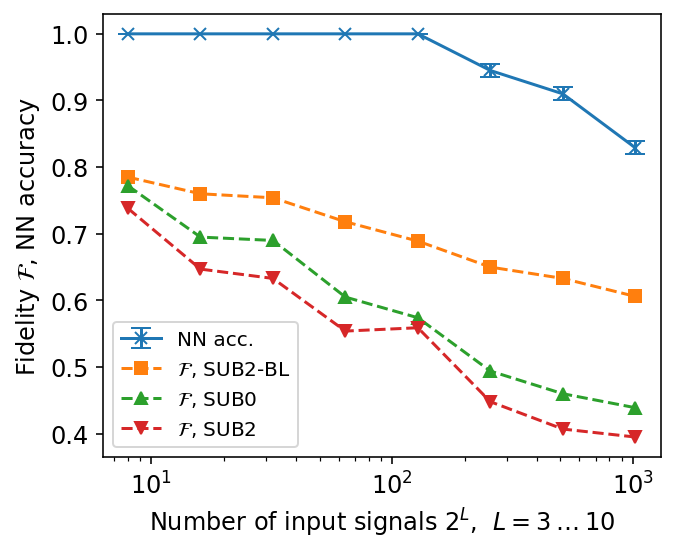}
 \caption{\label{fig-binwordsff} Fidelity $\mathcal{F}$ and neural-net (NN) recognition accuracy for sets of binary-word signals of size $S=2^L$ varying from 8 to 1,024.
 	The fidelity is plotted for three different ways of processing the measured memory signals, SUB0 (subtraction of the response to the zero input signal $I_0(u)$), SUB2 (subtraction of the response to a second readout compression cycle), and SUB2-BL (SUB2 with a linear baseline subtracted to make the signal zero at the endpoints $u=0,1$).
 	The SUB2-BL processed signals were also used as inputs to the NN, and for $p_\textit{sig}^{(s)}(u)$ plotted in Fig.~\ref{fig-bwmem}.
 	Error bars on the NN accuracy are from training the model multiple times.}
 \end{figure}

\subsection{\label{depend-grainprops}Dependence on grain material properties}
	The simulation parameters used in this work correspond to mm-scale rubber-like grains, and it is natural to wonder if the waveform memory effect would be observable in more sand-like media with smaller, harder grains.
	Two factors make this difficult to explore with available numerical resources:

(a) The sample compression $\delta_0$ (typically 0.05 in the present work) would need to be much smaller to avoid exceeding the yield strain of sand-like grains.
	As Fig.~\ref{fig-ffdelta0oo} shows, this in turn would require smaller signal-input amplitude $\gamma_0$ and would also reduce the number of contacts formed per grain upon compression (increase of $Z$ in Fig.~\ref{fig-thingsvst}).
	The combined effect would be to require many more grains than we are currently able to simulate ($\sim 10^5$).
	
(b) The characteristic time $t_c = R(\rho/E)^{1/2}$ would become shorter, requiring smaller simulation time steps and thus more steps for a quasistatic (small inertial number) simulation.

\section{\label{limits}Limits to memory complexity}
	The results of Sec.~\ref{demonstration} all used the suite of six input waveforms shown in Fig.~\ref{fig-inputs}(a) and also used a separate zero-input simulation starting from the same microscopic grain configuration for baseline subtraction, which would not be possible in physical experiments.
	In this section we introduce a physically-implementable baseline subtraction procedure (which requires longer simulations but turns out to have relatively little effect on the results) and also introduce systematically more complex input waveforms to explore the complexity limits of the waveform memory effect.
	
	For a more realistic baseline subtraction procedure, the simulations were extended to include an additional compression/decompression cycle after the readout decompression shown in Fig.~\ref{fig-thingsvst}.
	The shear stress $p_x - p_y$ measured during the final decompression was subtracted as a baseline from the shear stress measured during the first readout decompression, and the separate simulation with the zero input signal $I_0(u)$ was not used.
	This procedure  apparently works because the first readout decompression erases the stored memory by opening the frictional contacts that were made when the memory was stored, Fig.~\ref{fig-explanation}.
	As shown in Fig.~\ref{fig-binwordsff}, the measured memory fidelity $\mathcal{F}$ is only slightly reduced by switching to this baseline subtraction procedure.
	
	To create systematically more complex input signals, we have used shear inputs $I_s(u)$ formed from half-cycle cosine peaks as in Fig.~\ref{fig-inputs}(a) but now arranged so the integrated signals $J_s(u)$ follow $L$-bit binary words, giving a suite of $S = 2^L$ input signals for any chosen $L$ value (Fig.~\ref{fig-inputs}(b) and Appendix~\ref{app-inputsigs}).
	By increasing $L$, input signals with arbitrarily fine detail can be created.
	
	These binary-word input waveforms are constructed so $J_s(u)$ goes to zero at the endpoints $u=0,1$, Fig~\ref{fig-inputs}(b).
	This makes it possible to subtract a linear background between $u=0$ and $u=1$ from the recalled signals, which is found to significantly increase the fidelity $\mathcal{F}$ (Fig.~\ref{fig-binwordsff}).
	Note that both transformations used for the recalled signals -- subtraction of the granular-pack asymmetry background inferred from a second decompression followed by subtraction of a linear background to bring the endpoints to zero -- would be possible in experiments on a physical system.

	The waveform memory effect appears significant, if increasingly inaccurate, with $L$ as large as $10$  (the largest $L$ that was feasible with our computational resources, requiring 1,024 separate simulation runs), Fig.~\ref{fig-bwmem}.
	This means granular packs of the size simulated here ($10^4$ grains) have some limited ability to recall 1,024 different input waveforms.
	
	Some of the decrease in $\mathcal{F}$ with increasing word length $L$ appears to be due to modest amplitude and $u$-shifts of the recalled signals, so for an alternative measure of the memory capability a small neural net (NN) was trained to recognize the input signals in the presence of such shifts (Appendix~\ref{app-ml}).
	Then the accuracy with which the NN could recognize the recalled signals was measured. 
	The NN was 100\% accurate in classifying recalled signals for $L\leq 7$ (up to 128 distinct signals), with the accuracy falling off to 91\% for the 512 distinct $L=9$ signals,  and to 83\% for the 1,024 distinct $L=10$ signals (Fig.~\ref{fig-binwordsff}).

\section{\label{discuss}Discussion}

\subsection{\label{relationtohysteron}Relation to hysteron-based memory}
	Is there any connection between the waveform memory discussed in this work, and the hysteron-based memory~\cite{keim20, bense21, shohat22, lindeman25, lindeman25a, paulsen25} that has been seen in cyclically driven non-colloidal suspensions, glassy systems, and corrugated and crumpled sheets?
	Both sets of phenomena occur in random media containing a finite density of meso-scale hysteretic variables -- in the present case frictional contacts, demonstrated by loops in the (displacement - force) plane for a contact (Fig.~\ref{fig-m1m2h}).
	
	Unlike most of the hysteron-based memory studies, the present work used two orthogonal strain modes (compression and shear) and formed memories in a single cycle of the input variables.
	However Lindeman~\cite{lindeman25a} applied two distinct strain modes to a low-friction granular system and observed hysteron-based memory that depended upon the order in which the two types of strain were applied, similar to the way in which the waveform memory described here depends upon the order in which compression and shear are applied.
	And as described in Secs.~\ref{form-sample}, \ref{mem-expts} we trained granular packs with several cycles of compression before appying both strain modes together to encode the waveform memory.
	
	Perhaps a more significant difference between the present work and these earlier studies is that here one type of strain (compression) is applied at large amplitude as a reference signal, while a different type of strain (shear) is applied at low amplitude as signal input (Fig.~\ref{fig-compressmem}).
	This enables a large number of hysteretic variables (here frictional contacts) to be accessed independently of the amplitude of the memory input signal.

\subsection{\label{connectionwithelasticity}Possible connection with bulk elasticity}
	It was found above that the recalled memory signal $p_\textit{sig}$ (a shear stress) is proportional to the product of the compressive strain $\delta_0$ and the shear strain magnitude $\gamma_0$ with proportionality factor $\mathcal{G}E$, see Eqs.~\ref{psig},~\ref{ggdef}.
	On this basis $\mathcal{G}E$ might be regarded as a type of second-order elastic constant, relating stress to strain squared~\cite{norris97}.
	This viewpoint is bolstered by the weak dependence of $\mathcal{G}$ on system size, Fig.~\ref{fig-ggdelta0ng}(b).
	
	However attempting to cast the waveform memory as an elastic effect encounters several problems or at least complications:
	(i)~The bilinearity of  $p_\textit{sig}$ in the two strains $\delta_0$, $\gamma_0$ depends not only on both strains being small, but also on the ordering $\gamma_0 \ll \delta_0$ (Fig.~\ref{fig-ggdelta0ng}(a)).
	(ii)~The stress $p_\textit{sig}$ depends on the \emph{order} in which the strains $\delta_0$, $\gamma_0$ are applied (Fig.~\ref{fig-explanation}), making $\mathcal{G}E$ something like a non-abelian elastic constant~\cite{lindeman25a}.
	(iii)~The internal state of the stress of the granular medium must be general enough to encode a waveform, which would seem to go beyond ordinary elasticity theory.

\subsection{\label{experiments}Observability in experiments and with more sand-like grains}
	It should be possible to test the memory effect discussed here experimentally, using for example a pack of millimeter-scale rubber granules~\footnote{For experiments it would not be necessary for the grains to have precisely the multi-spherical shape shown in Fig.~\ref{fig-tetras}, which was used in the simulations for numerical efficiency.} confined in an apparatus (such as a rheometer) capable of smoothly applying compressive and shear strains while sensitively measuring the resulting stresses.
	
	It would be interesting to see if the waveform memory effect is observable in more sand-like media with harder grains.
	As discussed in Sec.~\ref{depend-grainprops} it is difficult to simulate waveform memory experiments using hard grains, but experiments using such grains should be feasible.
	
	The force laws used above essentially assume grain-scale elasticity coupled with more microscopic Coulomb-law friction at contacting surfaces~\cite{johnson85, johnson97}.
	This should be valid for large elastomer grains, but harder, more sand-like granular media might not have this separation of scales.
	Recent work has explored microscopic, asperity-based models of friction for jammed granular media~\cite{papanikolaou13,ikeda20}, and it would be interesting to see if the memory effect reported here persists for such models.

\subsection{\label{othermedia}Generalization to other types of random media}
	According to the explanation proposed in Fig.~\ref{fig-explanation}, the key property of granular packs that enables waveform memory is the formation of progressively more internal contacts as the system is compressed.
	This suggests that similar waveform memory should be observable in other systems such as fiber nests~\cite{bhosale22, gey25}, fiber bundles and yarns~\cite{panaitescu18, seguin22, dawadi25}, textiles~\cite{duhovic06, poincloux18, crassous24, singal24, gonzalez25}, and crumpled sheets~\cite{matan02, cambou11, lahini17, vanbruggen19}.
	
	It is also possible that the types of matter in which hysteron-based memory has been demonstrated~\cite{fiocco14, mukherji19,keim20, bense21, dawadi24, lindeman25a} might display similar waveform memory if subjected to two simultaneous strains~\cite{lindeman25a} as in  Fig.~\ref{fig-compressmem}: a large-amplitude strain as a reference input, along with a small-amplitude strain as a signal input.

\begin{acknowledgments}
	This work was completed in part with resources provided by University of Massachusetts Research Computing and Data and the Massachusetts Green High Performance Computing Center (GHPCC).
\end{acknowledgments}

\section*{Author Contributions}
	D.C. conceptualized this work, and wrote the draft manuscript.
	E.D. and D.C. jointly wrote the software used, carried out the investigation, and edited the manuscript.

\section*{Data Availability}
	All data plotted as points in the figures are openly available~\footnote{Data plotted as points in the figures are available in tabular form at \url{https://github.com/doncandela/gmem-data}.}.
	Due to their sizes, the datasets consisting of waveform data are not posted but they are available upon reasonable request to the authors.

\appendix

\section{\label{app-inputsigs} Input signals used for memory experiments}
	Each input signal $I_s(u)$ used in this work is the sum of $B^{(s)}$ non-overlapping half-cycle cosine bumps of width $\delta u$, center locations $u_{0b}^{(s)}$, and amplitudes $I_{0b}^{(s)}$,  see Fig.~\ref{fig-inputs}:
\begin{align}
	I_s(u) &= \sum_{b=0}^{B^{(s)}-1} I_{0b}^{(s)} f((u-u_{0b}^{(s)})/\delta u) \\
	\mbox{with~~} f(x) &=   \left\{ \begin{array}{l l}
		\cos(\pi x) & |x|<\tfrac{1}{2},\\
		0 & \mbox{otherwise.}
	\end{array} \right.  \nonumber
\end{align}
The integrated signal $J_s(u) = \int_0^u I_s(v)dv$ is therefore
\begin{align}
	J_s(u) &= \sum_{b=0}^{B^{(s)}-1} I_{0b}^{(s)} g((u-u_{0b}^{(s)})/\delta u) \\
	\mbox{with~~} g(x) &=   \left\{ \begin{array}{l l}
	         0 & x \leq -\tfrac{1}{2},\\
		(\delta u/\pi)(\sin(\pi x) + 1) & |x|<\tfrac{1}{2},\\
		2\delta u/\pi & x \geq \tfrac{1}{2}.
	\end{array} \right.  \nonumber
\end{align}
	Every set of input signals used included the zero signal  $I_0(u)=0$ along with $S$ non-zero signals, $s=1\dots S$.
	For each of the non-zero signals $\max_b|I_0^{(b)}| = 1$ so the maximum shear strain applied to the granular sample was $\gamma_0$ (typically $10^{-3}$).
	
\subsection{Initial set of signals.}
	These are the set of $S=6$ non-zero signals used in Sec.~\ref{demonstration}.
	These input signals used $\delta u = 0.2$ and the $u_{0b}^{(s)}$ and $I_{0b}^{(s)}$ values shown in Table~\ref{tab-sixinputs}.

\subsection{Binary-word signals.}
	These are the sets of $S=2^L$ non-zero signals based on $L$-bit binary words used in Sec.~\ref{limits}.
	These signals used a parameter $p=2.0$ setting the duration (in $u$) of the initial and final segments with $I_s(u)=0$ relative to the duration used for one bit $u_b$, therefore $u_b  = 1/(L+2p)$.
	Furthermore $\delta u = 1.0\times u_b$ was used.
	
	To compute input signal $I_s(u)$ or its integral $J_s(u)$, the bits $b_0\dots b_{L-1} = 0,1$ in the $L$-bit binary representation of $s-1$ are found, with $b_0$ the most significant bit.
	Then $I_s(u)$ has $B^{(s)} = L+1$ half-cosine bumps, with $L-1$ of the bumps giving the transitions between successive bits,
\begin{multline}
	I_{0b}^{(s)} = b_{b+1}-b_b,\ \  u_{0b}^{(s)} = (p+b+1)u_b \\
	\mbox{for~} b=0\dots(L-2).
\end{multline}
	The remaining two bumps give the transitions from and to the $I_s(u)=0$ periods near $u=0,1$,
\begin{align}
	I_{0(L-1)}^{(s)} &= b_0 - \tfrac{1}{2},\ \  u_{0(L-1)}^{(s)} = p u_b ,\\
	I_{0L}^{(s)} &=  \tfrac{1}{2} - b_{L-1},\ \  u_{0L}^{(s)} = (p+L)u_b. \nonumber
\end{align}

\begin{table}
\caption{\label{tab-sixinputs} Coefficients used for the initial set of input signals.}
\begin{ruledtabular}
\begin{tabular}{lll}
$s$ & $u_{0b}^{(s)}$ values & $I_{0b}^{(s)}$ values\\
\hline
 1 & [0.3] & [1.0]\\
 2 & [0.7] & [1.0]\\
 3 & [0.3, 0.7] & [1.0, 1.0]\\
 4 & [0.3, 0.7] & [--1.0, 1.0]\\
 5 & [0.3, 0.7] & [1.0, --1.0]\\
 6 & [0.2, 0.5, 0.8] & [0.5, --1.0, 0.5]\\
\end{tabular}
\end{ruledtabular}
\end{table}

\section{\label{app-dotprod} Dot-product evaluation of memory gain and fidelity}
	The dimensionless gain $\mathcal{G}$ and fidelity $\mathcal{F}$ for the memory effect for a set of $S$ integrated memory inputs ${J_s(u)}$, $s=1\dots S$ are calculated from generalized dot products between sets of signals.

	The value of the gain $\mathcal{G}$ was determined by minimizing squared error in Eq.~\ref{ggdef} integrated over $u$ and summed over the $S$ input signals,
\begin{equation}\label{eesq}
	\mathcal{E}^2 = \sum_{s=1}^S \int_0^1 \left( p_\textit{sig}^{(s)}(u) - \mathcal{G}E \delta_0\gamma_0 J_s(u) \right)^2  du.
\end{equation}
	Considering a set of $S$ $u$-dependent functions to be a  vector and defining the dot product
\begin{equation}
	\{A_s(u)\}_{s=1\dots S} \leftrightarrow \mathbf{A},\ \ \ \mathbf{A}\cdot\mathbf{B} = 
	 \sum_{s=1}^S \int_0^1 A_s(u) B_s(u) \, du
\end{equation}
the total error squared is
\begin{equation}\label{eesq2}
	\mathcal{E}^2 = (\mathbf{p}_\textit{sig} - \mathcal{G}E\delta_0\gamma_0\mathbf{J}) \cdot
	(\mathbf{p}_\textit{sig} - \mathcal{G}E\delta_0\gamma_0\mathbf{J}).
\end{equation}
	Minimizing $\mathcal{E}^2$ by setting $d\mathcal{E}^2/d\mathcal{G} = 0$ yields
\begin{equation}\label{gg}
	\mathcal{G} = \frac{1}{E\delta_0\gamma_0} \frac{\mathbf{J} \cdot \mathbf{p}_\textit{sig}} {\mathbf{J} \cdot \mathbf{J}}.
\end{equation}
	The dimensionless fidelity is defined as
\begin{equation}\label{ff}
	\mathcal{F} \equiv 1 -\left( \frac{\mathcal{E}^2}{\mathbf{p}_\textit{sig} \cdot \mathbf{p}_\textit{sig}} \right)^{1/2}
\end{equation}
which satisfies $0 \leq \mathcal{F} \leq 1$.
	If the recalled signals scaled by the optimized $\mathcal{G}$ perfectly match the input signals then the fidelity is $\mathcal{F} =1$, while conversely if the recalled signals are completely uncorrelated with the input signals $\mathcal{F} \ll 1$.

\section{\label{app-ml} Using a neural net to assess memory capability}
	The granular waveform memory studied here can be thought of as an imperfect  phonograph, i.e. analog waveform recording/playback system with compression of the granular pack $u \in (0,1)$ playing the role of time in an ordinary phonograph.
	The capability of a phonograph can be measured by the ability of listeners, untrained in its imperfections, to understand its output~\cite{sciam96}.
	Some transformations of the waveform such as small time or amplitude shifts or the addition of small-amplitude noise do not impair the recognition of the playback of a phonograph but do decrease the dot-product fidelity $\mathcal{F}$.
	
	To circumvent this limitation of $\mathcal{F}$, a neural net (NN) was trained to recognize the binary-word input signals described in Sec.~\ref{limits} with transformations as described above applied to them.
	Then it was measured how well the trained NN was able to classify the outputs of the granular-memory simulations.
	Note that the NN was trained independently of any granular memory simulations.
	
	PyTorch~\footnote{PyTorch, \url{https://pytorch.org/}} was used to implement  simple type of NN, a multilayer perceptron~\cite{zhang23}.
	It was found that this NN could be trained to greater than 99.9\% accuracy on the transformed input signals, so more complex NN's were not tried.
	The input signals $J_s(u)$ were discretized to 256 evenly-spaced $u$ values in $[0,1]$, and this was the size of the input layer of the NN.
		The input layer was followed by two 1024-element fully-connected hidden layers with ReLU activation, then a fully-connected output layer with $2^L$ elements for classification of the inputs according to the binary-word index $s$.
	
	The NN was trained with standard elementary methods~\cite{zhang23}: Kaiming-uniform weight initialization, cross-entropy loss function, stochastic gradient descent with batch size 32, no dropout.
	To provide sufficient training and validation data, 17,325 different transformations were applied to each of the $2^L$ input signals $J_s(u)$. 

	After the NN was trained, its accuracy was measured for classifying the $2^L$ recalled signals $p_\textit{sig}^{(s)}(u)$ generated by doing granular memory simulations (Fig.~\ref{fig-binwordsff}).
	Each input set (training and validation subsets of the training data, and recalled memories) was  normalized to make  the average across the set $\langle |J_s(u)|\rangle_{u,s} = 1$ before being used as input to the NN.

\section{\label{app-numerical} Numerical methods}

\subsection{\label{lammps} Correspondence with LAMMPS}
	Although lab-written DEM code was used for convenience in applying the complicated boundary motion used in this work, the numerical integration methods and contact force laws used (apart from model M2) correspond closely to those available in popular DEM software such as LAMMPS~\cite{lammps25}.

	Within LAMMPS there is an old style of granular force laws invoked with the \verb7pair_style gran/...7 commands, and a new, more flexible style invoked with \verb7pair_style granular7 followed by a \verb7pair_coeff7 command supplying additional parameters.
	For both old and new styles the default rolling and twisting torques are zero, as were used in this work.

	The repulsive-only damped Hertzian normal force law used in this work (Eq.~\ref{fn})
 corresponds to the old style command
 \begin{verbatim}
pair_style gran/hertz/history...limit_damping
 \end{verbatim}
 or the new style commands
 \begin{verbatim}
pair_style granular
pair_coeff * * hertz...limit_damping
 \end{verbatim}
 Here and below ellipses \verb7...7 indicate supplied values of force-law parameters.
	As in this work, the algorithm described by Luding~\cite{luding08} is used in LAMMPS to implement the Coulomb-law limit on transverse forces.

	The Hookean transverse force law used in this work with a Hertzian normal force law (model H, Eq.~\ref{fth}) is not available using old style commands, but it corresponds to the  new style commands
\begin{verbatim}
pair_style granular
pair_coeff * * hertz...&
   tangential linear_history...&
   limit_damping
 \end{verbatim}
 
	The simpler Mindlin transverse force law used in this work (model M1, Eq.~\ref{ftm1}) with rescaled force on contact unloading to avoid spurious energy creation is not available using old style commands, which do not include rescaling. Model M1 corresponds to the new style commands
\begin{verbatim}
pair_style granular
pair_coeff * * hertz...&
   tangential mindlin_rescale/force...&
   limit_damping
 \end{verbatim}

	The more accurate but numerically expensive Mindlin transverse force law used in this work (model M2) does not appear correspond to a currently-implemented LAMMPS force law.

 \begin{figure}
 \includegraphics[width=\linewidth]{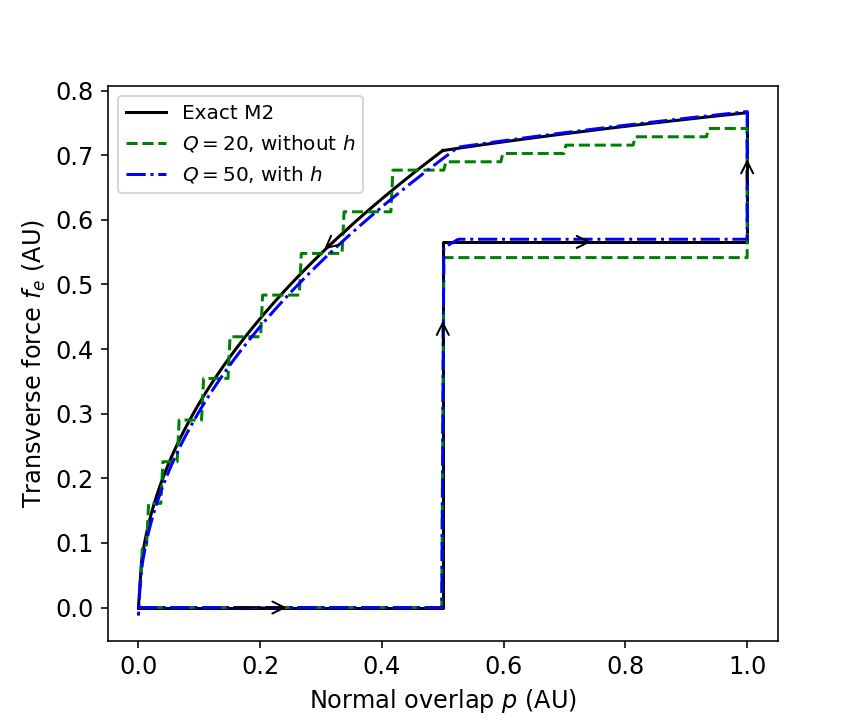}
 \caption{\label{fig-m2disc} Illustration of the discretized implementation of transverse force law M2, for the same grain-grain contact path shown in Fig.~\ref{fig-m1m2h}.
 	The solid black curve shows the exact  $f_e(p)$ from model M2, while the other curves  show $f_e(p)$ calculated using the discretized algorithm with $p_\mathit{max}=1.5$.
 	The dashed green curve shows the force calculated with a small number of slices $Q=20$ and omitting the term in $h$ in item~\ref{compfe} of App.~\ref{app-m2}.
 	Including the $h$ term smooths the steps in the computed $f_e(p)$ while increasing $Q$ increases the accuracy, as the blue dash-dot curve shows.
 }
 \end{figure}

\subsection{\label{app-m2}Implementation of model M2}
	In model M2, rather than storing a single elastic transverse vector force vector $\mathbf{f}_e$ as in model M1, a transverse-vector valued function $\mathbf{f}(q)$ is stored for each contact  (Sec.~\ref{depend-friction}) and transformed for joint motions of the grains at each simulation step~\cite{luding08, matuttis14}.
	Here $q=p^{1/2}$ with $p\geq0$ the grain-grain normal overlap, and conceptually $\mathbf{f}(q)$ is a continuous function on $q \in [0,\infty)$.
	As model M2 has not typically been used for DEM simulations, detailed steps for an implementation that was found to be usable are given here.

	The nominally continuous variable $q$ was discretized to $Q$ values evenly spaced in $[0,q_\mathit{max}]$ with $q_\mathit{max} = (p_\mathit{max})^{1/2}$ and  $p_\mathit{max}$ set larger than all (or nearly all) overlap values $p$ that occur over the course of the simulation~\footnote{For this work, as model M2 received limited use, $p_\mathit{max}$ was simply set manually to be slightly larger than the maximum $p$ value found in the DEM runs.}.
	Therefore the actual history variables stored for each contact are at set of $Q$ transverse vectors $\{\mathbf{f}_j\} \equiv \mathbf{f}(j\times dq)$, $j = 0\dots(Q-1)$ with $dq = q_\mathit{max}/(Q-1)$.
	For the results using model M2 shown in Fig.~\ref{fig-varymodel}, $Q = 100$ was used.
	
	At each time step in the simulation the following steps are carried out for each grain-grain contact to calculate the transverse force $\mathbf{f}_t$ transmitted by the contact and to update the history variables $\{\mathbf{f}_ 0\dots \mathbf{f}_{Q-1}\}$ based on the current normal overlap $p$, normal force $f_n$ and sliding velocity $\dot{\bm{\sigma}}$, and the sliding motion $\Delta\bm{\sigma}$ in this contact since the previous simulation step.
	Note $\mathbf{f}_t$,   $\{\mathbf{f}_ j\}$,  $\dot{\bm{\sigma}}$ and $\Delta\bm{\sigma}$ are all vectors transverse to the contact normal.
\begin{enumerate}
\item The following quantities are computed:
	\begin{enumerate}
	\item $j_p = \mbox{min}(\mbox{floor}[x_p],Q-2)$ with $x_p = p^{1/2}/dq$.
	Thus $j_p \in [0,1\dots(Q-2)]$, and $\mathbf{f}_j$ gives $\mathbf{f}(q)$ at the discretized $q$ point closest below $p^{1/2}$.
		
	\item $h = 1 + j_p - x_p$, so $h \in [0,1)$.
	\end{enumerate}

\item
	\begin{itemize}
	\item If the contact is new,   $\mathbf{f}_j \leftarrow \mathbf{0}$ for $j=0\dots(Q-1)$.
	(Conceptually $\mathbf{f}(q)  \leftarrow \mathbf{0}$ for $q \in [0,\infty)$.)

	\item Otherwise the $\{\mathbf{f}_j\}$ saved in the previous simulation step are retrieved and  corrected for joint tumbling and spinning motions of the grains~\cite{luding08, matuttis14}.
	Then the $\{\mathbf{f}_j\}$ are adjusted for the reduction in $p$ if any, by setting
	$\mathbf{f}_j \leftarrow \mathbf{0}$ for $j=(j_p+1)\dots(Q-1)$.
	(Conceptually $\mathbf{f}(q) \leftarrow \mathbf{0}$ for $q\in (p^{1/2},\infty)$.)
	\end{itemize}

\item $\mathbf{f}_j \leftarrow \mathbf{f}_j +k_M\Delta\bm{\sigma}$ for $j= 0\dots j_p$,  to adjust the $\{\mathbf{f}_j\}$ for sliding motion since the previous simulation step.
	This adds approximately  $k_Mp^{1/2} \Delta\bm{\sigma}$ to $\mathbf{f}_c$.
	(Conceptually $\mathbf{f}(q) \leftarrow \mathbf{f}(q)  +k_M\Delta\bm{\sigma}$ for $q \in (0,p^{1/2})$.)

\item The following quantities are computed:
	\begin{enumerate}
	\item\label{compfe}The transverse elastic force $\mathbf{f}_e= dq(\tfrac{1}{2}\mathbf{f}_0 + \mathbf{f}_1 + \dots + (1-h)\mathbf{f}_{j_p} + \dots \mathbf{f}_{Q-1})$, equivalently $\mathbf{f}_e = dq((\sum_{j=0}^{Q-1} \mathbf{f}_j) - \tfrac{1}{2}\mathbf{f}_0 - h\mathbf{f}_{j_p})$ \footnote{The
term in $h$ is added so that $\mathbf{f}_e$ does not change discontinuously when $j_p$ decrements as $p$ decreases on contact unloading, see Fig.~\ref{fig-m2disc}. Such discontinuous changes in contact forces tend to destabilize dense granular packs, and including the $h$ term was found to enable stable simulations using model M2 with smaller $Q$ values, saving computational resources.}.
	(Conceptually $\mathbf{f}_e= \int_0^\infty \mathbf{f}(q)dq$.).
	
	\item The Coulomb-limit transverse force magnitude $f_c = \max(\mu f_n,0)$.

	\item The transverse damping force $\mathbf{f}_d =  \gamma_M p^{1/2}\dot{\bm{\sigma}}$.

	\item The trial transverse force $\mathbf{f}_t^0 = \mathbf{f}_e + \mathbf{f}_d$ and its magnitude $|\mathbf{f}_t^0|$.
	\end{enumerate}
\item
	\begin{itemize}
	\item If $|\mathbf{f}_t^0| \leq f_c$, the contact is below the sliding limit, so $\mathbf{f}_t = \mathbf{f}_t^0$.

	\item Otherwise the contact is sliding, so $\mathbf{f}_t$ is computed and the history variables $\{\mathbf{f}_j\}$ are modified as follows, analogously to the algorithm of Ref.~\cite{luding08}: 
		\begin{enumerate}
		\item $\mathbf{f}_t = (f_c/|\mathbf{f}_t^0|)\mathbf{f}_t^0.$
		
		\item $\Delta\mathbf{f}_e = \mathbf{f}_t -\mathbf{f}_t^0.$

		\item $\mathbf{f}_j \leftarrow \mathbf{f}_j + \Delta\mathbf{f}_e/(dq(j_p+\tfrac{1}{2} -h))$ for $j = 0\dots j_p$ . 
		This adds precisely $\Delta\mathbf{f}_e$ to $\mathbf{f}_e$ computed as in \ref{compfe} above.
		(Conceptually $\mathbf{f}(q) \leftarrow \mathbf{f}(q)  +\Delta\mathbf{f}_e/p^{1/2}$ for $q \in (0,p^{1/2})$.)
		\end{enumerate}
	These steps leave $|\mathbf{f}_t| = f_c$ and the next-step trial force will be equal to $\mathbf{f}_t$ in the absence of additional sliding $\Delta\bm{\sigma}$ or changes in the sliding rate $\dot{\bm{\sigma}}$ .
	\end{itemize}
\end{enumerate}
Figure ~\ref{fig-m2disc} shows an example of this discretized implementation of model M2.

\bibliography{gmp1bib,gmbib}

\end{document}